\documentclass[%
 reprint,
superscriptaddress,
usenames,dvipsnames
preprintnumbers,
 amsmath,amssymb,
 aps,
]{revtex4-2}

\usepackage[normalem]{ulem}
\usepackage[colorlinks=true, allcolors=blue]{hyperref}
\usepackage{academicons}
\usepackage{xcolor}
\usepackage{orcidlink}
\usepackage{flushend}

\usepackage{graphicx}
\usepackage{dcolumn}
\usepackage{bm}
\usepackage{hyperref}
\usepackage{color}
\usepackage{multirow}

\begin{document}

\begin{flushright}
    BONN-TH-2025-12
\end{flushright}

\title{From obstacle to opportunity: uncovering the silver lining of pileup}

\author{Biplob Bhattacherjee\,\orcidlink{0000-0003-3668-8305}}
\email{biplob@iisc.ac.in}
\affiliation{Centre for High Energy Physics, Indian Institute of Science, Bangalore 560012, India}

\author{Abhinav Kumar\,\orcidlink{0009-0003-8060-8648}} 
\email{abhinavk23@iitk.ac.in}
\affiliation{Department of Physics, Indian Institute of Technology, Kanpur 208016, India} 

\author{Swagata Mukherjee\,\orcidlink{0000-0001-6341-9982}} 
\email{swagata@iitk.ac.in}
\affiliation{Department of Physics, Indian Institute of Technology, Kanpur 208016, India} 

\author{\\Rhitaja Sengupta\,\orcidlink{0000-0003-2293-8684}} 
\email{rsengupt@uni-bonn.de}
\affiliation{Bethe Center for Theoretical Physics and Physikalisches Institut der Universit\"at Bonn, \\
Nu\ss allee 12, Bonn 53115, Germany}

\author{Anand Sharma\,\orcidlink{0009-0005-5015-7596}}  
\email{anandsharma2@iisc.ac.in}
\affiliation{Centre for High Energy Physics, Indian Institute of Science, Bangalore 560012, India}


\begin{abstract}
\vspace{0.3cm}
The lack of evidence for Beyond Standard Model (BSM) particles might be due to their light mass and very weak interactions, as exemplified by  BSM long-lived particles (LLPs). Such particles can be produced from $B$ or $D$ hadron decays. Typically, the high values of pileup (PU) in hadron colliders are expected to pose a major challenge in light new physics searches. We propose a fresh perspective that counters this conventional wisdom: instead of viewing PU solely as an impediment, we highlight its potential benefits in searches for light LLPs from $B$ or $D$ hadron decays at HL-LHC and FCC-hh. In particular, certain forward detectors in LHC experiments, such as the Zero Degree Calorimeters (ZDC), which are currently not utilized for LLP searches, can be repurposed with strategic modifications to play a crucial role in this endeavor. 
Leveraging a combination of forward and central detectors, along with smart strategies for triggering and offline analysis, we demonstrate the potential for exploring light LLPs in high PU scenarios.
\end{abstract}

\maketitle

Typical hadron colliders, commonly thought of as discovery machines, operate at very high collision rates to maximize their discovery potential. Consequently, multiple proton-proton ($pp$) interactions occur simultaneously, whenever there is a bunch-crossing. The interaction involving the largest momentum exchange between the protons is called the ``hard-scatter'' interaction, and the remaining interactions are collectively known as ``pileup'' (PU). Generally, most PU interactions are low-energy collisions which contaminate the high-energy collision of interest. Thus, PU, which are nothing but soft QCD processes, pose a significant challenge during high-luminosity runs of hadron colliders, where they can obscure the primary interaction. It was found in past studies that PU interactions significantly degrade the resolutions of missing transverse energy~\cite{CMS-PAS-JME-12-002}, transverse momentum ($p_T$) of jets~\cite{CMS:PU_13TeV}, 
and jet mass, which are crucial variables employed in suppressing the QCD background in numerous Beyond Standard Model (BSM) searches. To combat PU, various mitigation strategies are used. At the Large Hadron Collider (LHC), the CMS experiment initially employed charged-hadron subtraction to reduce the impact of PU on jet reconstruction. This was later superseded by the more advanced pileup-per-particle identification (PUPPI) technique~\cite{PUPPI}. 

Traditionally, searches for BSM physics in hadron colliders have focused on TeV-scale particles with strong couplings, seeking high-$p_T$ signatures (for eg.~\cite{Swagata_HighMass_BSM}).
However, as time passed and hints of TeV-scale BSM particles remained elusive, a complementary class of lighter and more weakly coupled particles has gained interest. 
Due to their weak couplings, they are often long-lived (for eg.~\cite{LLP_review, LLP_backward_moving, LLP_L1trigger_HLLHC}), traversing significant distances without interacting before ultimately decaying to Standard Model (SM) particles.
This shift in interest to light new physics has now increased the importance of low-energy collisions.
Initiatives like CMS's data scouting program at the high level trigger (HLT)~\cite{CMS_HLT,Mukherjee_scouting_parking} and at the L1 trigger~\cite{CMS_L1T,L1_scouting_CMS}, and the data parking program~\cite{CMS_scouting_parking}, demonstrate the increasing recognition of low-energy collisions' value.

In BSM searches, PU is viewed as a nuisance due to its detrimental impact on trigger rate and on performance of object reconstruction. 
The influence of PU is even more pronounced for searches involving low-mass 
BSM particles, as the energy scales 
in these searches often overlap with those from PU interactions.
As the PU levels increase from the current LHC operating at a center-of-mass energy of $\sqrt{s} = 13.6 \, \text{TeV}$ (with a PU of approximately 40--60) to the High-Luminosity (HL) LHC (at $\sqrt{s} = 14 \, \text{TeV}$, with PU $\sim$ 140--200), and eventually to future hadron colliders like the FCC-hh (at $\sqrt{s} = 100 \, \text{TeV}$, with PU projected between 500--1000)~\cite{FCC_hh_1,FCC_hh_2}, the challenges in searching for light new physics will become progressively more daunting.
On the other hand, the huge amount of PU increases the statistics of low-energy collisions.
This raises an intriguing question: if BSM physics lies in the low-mass region, accessible through low-energy collisions, can PU be leveraged as a tool for new physics discovery, rather than merely being a hindrance? 

In the present work, we explore if there is any unforeseen advantage of PU for light weakly coupled new physics. We use the physics case of long-lived particles (LLPs) to illustrate this idea.
LLPs arise in various BSM scenarios, and are often produced through the decay of $B$ or $D$ hadrons~\cite{Feng:2017vli,Kling:2018wct}, which form during the hadronization of $b$ or $c$ quarks, respectively. The production of $B$ and $D$ hadrons is expected to increase significantly with the evolution of hadron colliders.
The only factor usually considered to contribute to this enhancement in $B$ and $D$ production is the increasing $\sqrt{s}$ for future colliders.
The impact of increasing $\sqrt{s}$ from 14~TeV to 100~TeV is noteworthy. 
To quantify this, we used \texttt{PYTHIA 8}~\cite{pythia6,pythia8} event generation software with 
the \texttt{Tune:pp$=$14}~\cite{pythia_tune} parameter tune and the NNPDF2.3 LO parton distribution function (PDF) set~\cite{NNPDF}. With this setup, we found that while the total soft QCD cross section increases moderately from 101~mb to 139~mb, the $D$ hadron production cross section increases almost two-fold, 
and the $B$ hadron production cross section increases five-fold, 
as we go from 14~TeV to 100~TeV. 

Another factor is the increased integrated luminosity, which is expected to increase 10-fold in going from the HL-LHC to FCC-hh.
This, in turn, increases the average number of pile-up vertices per bunch crossing, 
contributing to an increased number of $B$ and $D$ hadrons, which is due to the rise in PU interactions.
Although this effect is included in the integrated luminosity, what is often overlooked is the fact that even if the selected primary vertex of an event doesn't have associated light new physics, other vertices in that bunch crossing can have $B$ or $D$ hadrons, contributing to production of light LLPs. To check the effect of PU on $B$ and $D$ production rate, we conducted a study at a center-of-mass energy $\sqrt{s} = 14$~TeV with all soft QCD processes enabled, multi-parton interactions switched on, and $\text{PU}=0$. Using $10^8$ such $pp$ collision events, we obtained the total $pp$ cross-section to be $\sim$101~mb. Interestingly, 
$\sim$0.9\% of the events have at least 1 $b\bar{b}$ pair, while around 10\% of the events have at least 1 $c\bar{c}$ pair. Extrapolating to $\text{PU}=200$, in HL-LHC, we estimate that approximately 2-4 $B$ hadrons and $\sim$40 $D$ hadrons
will be produced per $pp$ bunch-crossing. This estimate is based on the following scaling relation: 
\begin{equation}
    n_{b/c} ({\rm PU} \simeq n_{\rm PU})= 2\times N_{b/c} ({\rm PU} \simeq 0) \times \frac{n_{\rm PU}}{n_{\text{event}}^{\text{SoftQCD}}}
    \label{eq:scaling}
\end{equation}
where $n_{\rm PU}=200$, which is the highest average PU expected during HL-LHC; $n_{b/c} ({\rm PU} \simeq n_{\rm PU})$ is the number of $B$ (or $D$) hadrons
expected to be produced in each $pp$ bunch-crossing with ${\rm PU} \simeq n_{\rm PU}$; $N_{b/c} ({\rm PU} \simeq 0)$ is the total number of $b\bar{b}$ (or $c\bar{c}$) pairs produced in $n_{\text{event}}^{\text{SoftQCD}}$ number of soft QCD $pp$ collision events generated with ${\rm PU} \simeq 0$. 
The factor of $2$ in Eq.~(\ref{eq:scaling}) comes from each $b\bar{b}$ or $c\bar{c}$ pair in a bunch crossing hadronizing into two $B$ or $D$ hadrons.
A similar study at $\sqrt{s} = 100$~TeV yields even more promising results. Here we found that, with $\text{PU}=500$, in FCC-hh, we expect $\sim$40 $B$ hadrons and $\sim$200 $D$ hadrons 
to be produced per $pp$ bunch-crossing. A significant fraction of $B$ and $D$ hadrons produced in PU interactions will travel towards the forward direction (see the pseudorapidity ($\eta$) distributions in Appendix\,\ref{appA_pt_pz_plot}, Fig.~\ref{fig:Eta_had_both}). Consequently, a sizable number of BSM LLPs may be produced in the forward region.
However, 
a non-negligible fraction of the $B$ or $D$ hadrons will travel towards the more centrally located barrel and the endcaps of the general purpose experiments. In this paper, we will point out the importance and complementary capabilities of two types of detectors: general purpose main experiments surrounding the $pp$ interaction point (IP), and dedicated forward detectors covering the high--$\eta$ region, in the context of HL-LHC and FCC-hh. 


We explore dark Higgs bosons as our first benchmark, featuring a scalar dark Higgs field~\cite{dark_higgs}, which is a singlet under the SM gauge group and couples to the SM sector only through its couplings with the SM Higgs field.  
The physical dark Higgs boson, denoted as $\phi$, couples to SM fermions via a mixing angle with the SM Higgs boson. We consider the production mode $B^{\pm}/B^0 \to \phi~ +~ K^{\pm}/K^0$, where $K^{\pm}$/$K^0$ are charged/neutral kaons. Depending on its mass, $\phi$ can decay into photons, muons, pions, kaons, or gluons. Assuming a very small mixing angle, $\phi$ can be long-lived.

As a second benchmark model, we consider Heavy Neutral Leptons (HNLs)\,\cite{Abdullahi:2022jlv}, 
denoted by $N$.
They can be produced from the decay of $B$ hadrons~\cite{CMS_HNL_BParking} which can happen in various ways such as, $B^{0} \to D^{\pm} l^{\mp} N$ or $B^{\pm} \to l^{\pm} N$ and so on. 
HNLs can be produced in two-body decays (e.g., $D^{\pm} \rightarrow \ell^{\pm} N$) or three-body decays (e.g., $D^{\pm} \rightarrow K^{0}\,\ell^{\pm} N$) of $D$ mesons~\cite{bondarenko2018phenomenology}. However, the strongest limits arise from the two-body channels, which benefit from larger branching ratios and greater phase-space availability. Consequently, in this work we focus on the two dominant production channels involving muons from $D$-meson decays, namely $D^{\pm} \rightarrow \mu^{\pm} N$ and $D_s^{\pm} \rightarrow \mu^{\pm} N$. Depending on the HNL mass, the HNL may subsequently decay either leptonically or semileptonically through two- or three-body decay modes.


We estimate the efficiencies of various sub-detectors to detect LLP decays within their decay volumes satisfying specified energy requirements. 
Here, the detector's efficiency is defined as the ratio of the number of LLPs decaying within a particular detector and having energy above some threshold to the total number of LLPs produced. 
While many studies present results in terms of upper limits on branching ratios or couplings, we provide geometrical efficiencies for a grid of LLP masses and lifetimes. This approach is more general and versatile, as it isolates the detector's performance from a specific physics model. By providing our results in this format, these efficiencies can be easily converted to bounds on branching ratios or couplings relevant to a desired theoretical model. This makes our findings broadly applicable to a variety of LLP scenarios.
The final analysis efficiency will comprise contributions also from triggering efficiency, object reconstruction efficiency, and additional analysis-level cuts aimed at background rejection. It is important to note that our efficiency calculations assume a single $B$ (or $D$) hadron per bunch-crossing, with the LLP originating from its decay. 
However, in reality, multiple $B$ or $D$ hadrons are produced in a single bunch-crossing, as previously discussed. Incorporating this enhancement would significantly improve the efficiency per bunch-crossing, leading to a more optimistic final analysis. Consequently, this would increase the discovery potential, or allow for more stringent limits to be placed on the phase space of interest.
Note that the efficiencies computed here are from the soft QCD processes, which contribute on top of the usually simulated hard QCD processes in most studies.

A dedicated forward detector, FOREHUNT~\cite{forehunt_paper}, has been proposed for the 100~TeV FCC-hh, with the primary goal of detecting light LLPs produced from the decay of hadrons. 
The detection efficiency of the FOREHUNT detector with configuration C for $\phi$ ($N$) can go up to 0.84\% (0.77\%) for a $m_\phi=4$\,GeV ($m_N=1.5$\,GeV) and a decay length of 0.1\,m with a 1\,TeV energy requirement on the LLP (see the efficiency grids given in Appendix\,\ref{appB_fms_zdc_faser_eff}, Fig.\,\ref{fig:forehunt_C_eff_E_1TeV}).  
Additionally, the general purpose detector in FCC-hh will feature a forward muon spectrometer (FMS), covering a pseudorapidity range of approximately $2.5 < |\eta| < 6.0$. Located between $z = \pm20.0$ meters and $z = \pm24.5$ meters, the FMS will make use of the forward dipole magnet to deflect the muon tracks. 
We find that the FMS has a maximum LLP detection efficiency of 2.1\% for $m_{\phi}=4$\,GeV and $c\tau=1$\,m after applying an energy requirement of $>$50\,GeV, while it reaches up to 1.7\% for $m_N \sim 1-1.5$\,GeV and $c\tau \sim 0.1$\,m (see the efficiency grids given in Appendix\,\ref{appB_fms_zdc_faser_eff}, Fig.\,\ref{fig:FMS_eff_E_50GeV}).
Since a detailed background analysis has not yet been performed, the energy cuts were chosen as initial estimates based on the expected energies of the LLPs. For instance, given that LLPs are expected to be highly energetic in the forward direction but less so in the transverse direction, a cut of at least 1\,TeV was applied for forward detectors at a 100\,TeV collider, while a 50\,GeV cut was used for the transverse detectors. This same logic was applied to the energy cuts for other detector configurations.
The reported efficiency numbers are conservative, as it is computed assuming only one $B$ (or $D$) hadron produced per bunch-crossing, while in reality, in FCC-hh, multiple $B$ (or $D$) hadrons will be produced per bunch-crossing. 
Since the PU vertices are all soft QCD interactions, the efficiency of detecting LLPs from $B$ (or $D$) hadron decays per bunch-crossing increases by a factor of around 40 (or 200) at the FCC-hh assuming a PU=500.

The production cross section of $B$ and $D$ hadrons at 100~TeV is currently plagued by large uncertainties, primarily due to the dominant contribution from gluons at very small Bjorken-$x$ values~\cite{FCC_physics}. Our present understanding of PDFs is limited in this regime~\cite{FCC_hh_SM_process}. About 50\% of $B$ hadrons produced at 100 TeV within the forward pseudorapidity range of $2.5 < |\eta| < 5$ are expected to come from gluons with momentum fraction $x \le 10^{-5}$. However, by the time FCC-hh commences operation, we can expect significant improvements in our knowledge of gluon PDFs at low-$x$. The forthcoming FCC-eh collider, scheduled to operate before FCC-hh, will enable precision measurements of gluon PDFs at small-$x$, achieving uncertainties of just a few percent down to $x \simeq 10^{-6}$. The pursuit of precise low-$x$ gluon PDF measurements will continue to advance through a concerted effort. Building on the foundation laid by the ongoing LHC program, future contributions from the HL-LHC, EIC, FCC-eh, and FCC-hh will further refine our understanding of low-$x$ gluon PDFs~\cite{pdf_LHC, pdf_EIC_1, pdf_EIC_2}.
%

It is evident that the searches for LLPs from $B$ or $D$ decays will benefit significantly at the FCC-hh, owing to both the increased $\sqrt{s}$ and higher PU. However, let us also explore the potential of these searches at the HL-LHC, where there will be a negligible increase in $\sqrt{s}$ w.r.t current LHC run, but the significantly increased PU will ironically enhance $B$ and $D$ production, providing a unique opportunity to leverage this phenomenon for LLP searches. We will explore the detection of LLPs in three distinct regimes. Firstly, we will discuss the ForwArd Search ExpeRiment (FASER)'s capabilities in detecting high-$c\tau$ LLPs. Next, we will examine the potential of general-purpose detectors, such as CMS, for detecting low-$c\tau$ LLPs. Finally, we will investigate options for detecting LLPs with intermediate $c\tau$ values. There are other existing and proposed sub-detectors and experiments~\cite{FACET_1,FACET_2, CODEX_b_1, MoEDAL_MAPP} that can facilitate LLP searches. Our goal is not to provide an exhaustive list of those. 

FASER~\cite{FASER_1,FASER_2,FASER_3} is a dedicated experiment searching for light, extremely weakly-interacting particles. Situated 480~meters downstream of the ATLAS IP~\cite{ATLAS_det_run3} with a cylindrical decay volume of 1.5~meters in length and 10~cm in radius, FASER is aligned with the collision axis and shielded by 100~meters thick layer of concrete and rock. 
The experiment features a multi-component detector for charged and neutral particles, with a cylindrical decay volume of 1.5~meters in length and 10~cm in radius. FASER covers the pseudorapidity range of $\eta > 8.8$.
FASER~\cite{FASER_1,FASER_2,FASER_3} is a dedicated experiment searching for light, extremely weakly-interacting particles. Situated 480~meters downstream of the ATLAS IP~\cite{ATLAS_det_run3}, FASER is aligned with the collision axis and shielded by 100~meters thick layer of concrete and rock. The experiment features a multi-component detector for charged and neutral particles, with a cylindrical decay volume of 1.5~meters in length and 10~cm in radius. FASER covers the pseudorapidity range of $\eta > 8.8$. 
For $\phi$ detection, applying an energy requirement of $>$500~GeV yields efficiencies in FASER up to $1.2\times10^{-5}$\% in the region of phase space where $m_{\phi}\sim 4$~GeV and $c\tau\sim1$~meter. For HNL detection, a similar energy requirement of $>$500~GeV results in efficiencies in FASER reaching up to $3.4\times10^{-5}$\% in the region of phase space where $m_N \sim 1.5$~GeV and $c\tau \sim 1$~meter, as shown in Fig.~\ref{fig:FASER_eff_E_500GeV} in Appendix~\ref{appB_fms_zdc_faser_eff}. 

So far, our discussion has focused on forward detectors located beyond $|\eta|=2.5$. However the region $|\eta|<2.5$ will also capture a non-negligible fraction of $B$ and $D$ hadrons, both in 14~TeV and in 100~TeV.
We, therefore, study the detection efficiencies for $\phi$ and $N$ in the muon spectrometer (MS) of the CMS detector at HL-LHC and the FASER detector (see Figs.~\ref{fig:FASER_eff_E_500GeV} and \ref{fig:CMS_eff_E_20GeV} in Appendix\,\ref{appB_fms_zdc_faser_eff}).
These two are sensitive to two different ends of the parameter space -- the CMS MS provides sensitivity to particles with low $c\tau$ as compared to FASER.
Here, we have applied a pseudorapidity cut of $|\eta| < 3$, consistent with the coverage of the CMS high-granularity calorimeter~\cite{CMS_hgcal}, which extends up to $|\eta| = 3$. 
These efficiency plots are presented in a 2D plane, with the LLP mass on the X-axis and the LLP $c\tau$ on the Y-axis. A minimum energy requirement of $E_{\text{LLP}}>20$~GeV is applied for both benchmark scenarios.
Clearly, the central region of general-purpose detectors can also significantly contribute to light LLP searches. 

In the context of HL-LHC, we've discussed detectors suitable for LLPs with high $c\tau$, such as FASER, located 480~meters from the IP. We've also explored detectors for LLPs with low $c\tau$, namely subdetectors of general-purpose detectors surrounding the IP, extending up to $\mathcal{O}(10)$~meters. 
An intermediate region, spanning from $\mathcal{O}(10)$\,m to 480\,m, corresponding to the region between the general-purpose detectors and far detectors, like FASER, remains largely unexplored. 
Remarkably, existing detectors in this region, although primarily used for other purposes, can be repurposed for LLP searches, offering great potential for detecting LLPs with intermediate $c\tau$ values.
The CMS experiment's two Zero Degree Calorimeters (ZDCs)~\cite{zdc_1,zdc_2}, located $\pm140$ meters from the IP, detect neutral particles at pseudorapidity $|\eta|>8.3$. Each ZDC consists of electromagnetic (19 radiation lengths) and hadronic (5.6 interaction lengths) sections, designed to measure forward neutral particles like neutrons and photons. Charged particles are typically deflected away by magnets before reaching the ZDCs. However, if a neutral LLP decays to charged SM particles just before the ZDC, the ZDC would detect a charged particle. If HL-LHC upgrades equip the ZDCs with two layers of tracking detectors before the calorimeter sections, observing a charged particle in the ZDC could serve as a distinctive signature for BSM LLPs. Furthermore, installing an additional two layers of tracking detectors after the ZDC would allow for the identification of the detected charged particle as a muon or not, providing valuable information for suppressing backgrounds from neutral LLPs in the SM, such as $K_{L}$ or neutrons.

\begin{figure}[hbt!]
\includegraphics[width=0.9\linewidth]{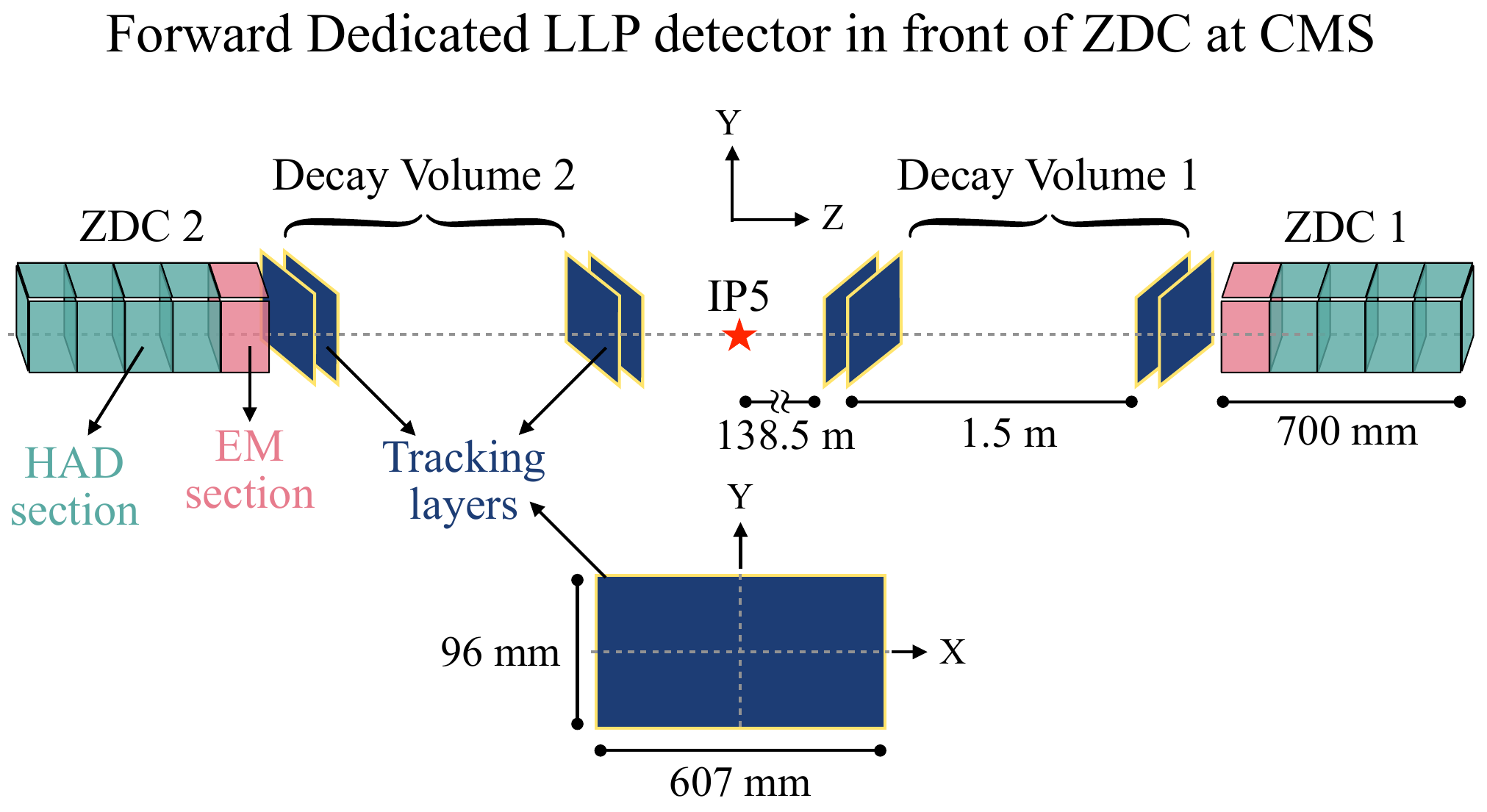}
\caption{\label{fig:zdc_det} Schematic view of the CMS ZDC detectors along with the proposed decay volumes marked by the tracking layers.}
\end{figure}

\begin{figure}[hbt!]
\includegraphics[width=0.8\linewidth]{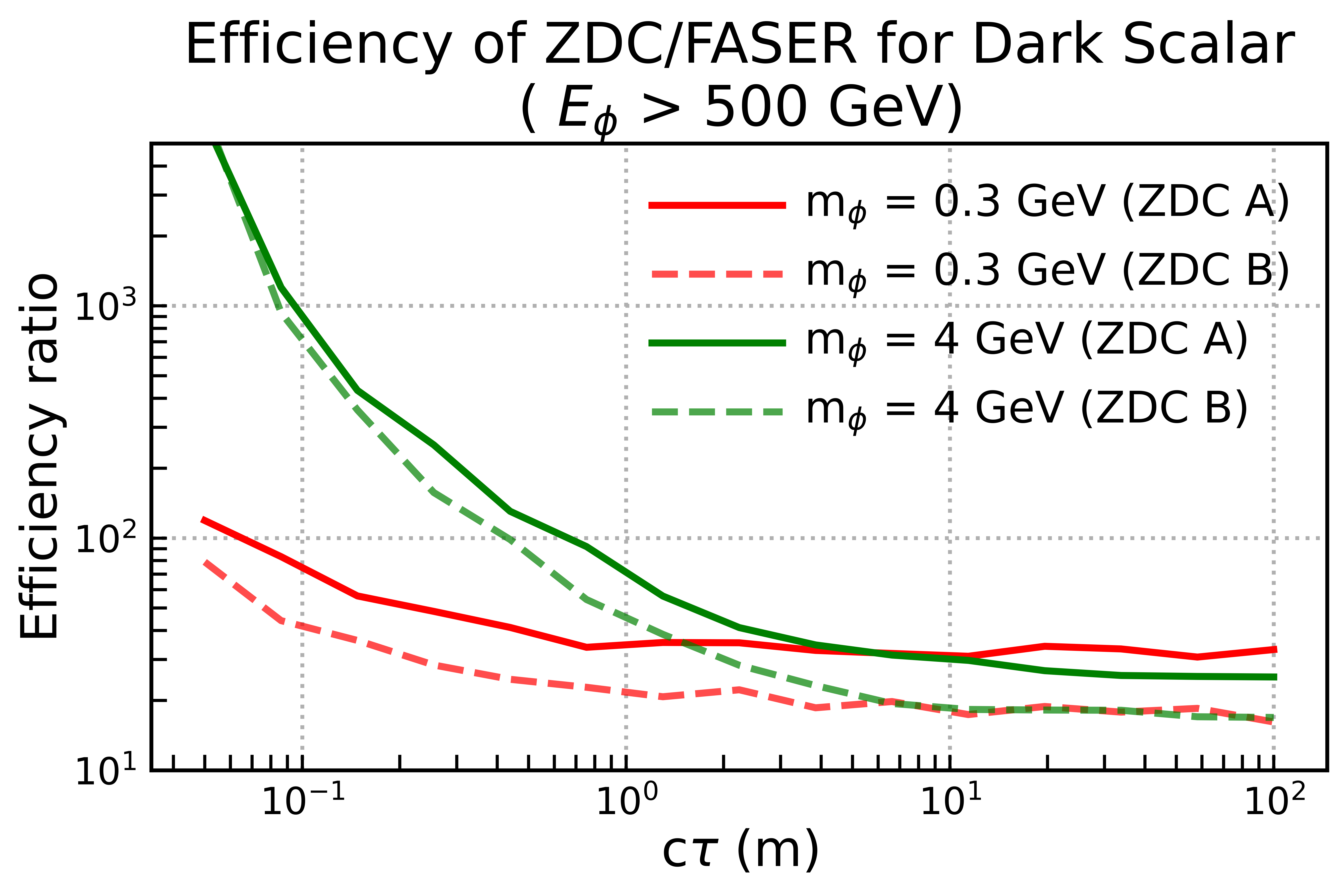}
\hskip 5 pt
\includegraphics[width=0.8\linewidth]{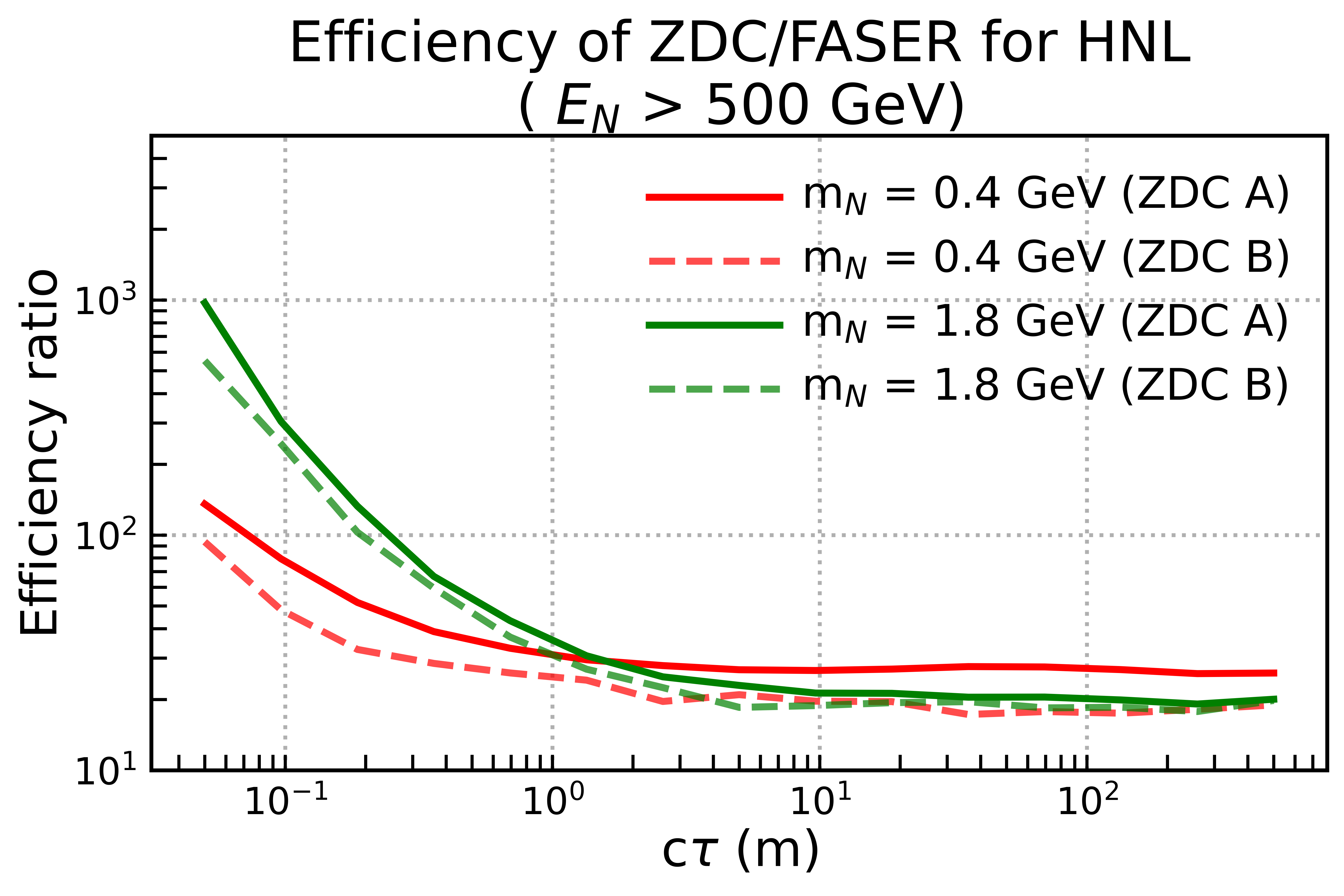}
\caption{\label{fig:zdc_vs_faser} Ratio of LLP detection efficiencies in ZDC and FASER. The ratio is calculated by dividing the efficiency of detecting LLP in the ZDC by the efficiency of detecting LLP in FASER, as a function of $c\tau$. The energy of LLP is required to be $>$500~GeV for both ZDC and FASER. The top plot is for $\phi$ and the bottom plot is for HNL.}
\end{figure}

To assess the capabilities of ZDC, we simulated a cuboid decay volume, denoted as ZDC A, with dimensions 1.5\,m in length, 96\,mm in height, and 607\,mm in width. The height and width were chosen to match the size of the box containing the ZDC~\cite{CMS_ZDC_dimension}, while the length was set to 1.5~meters, consistent with the length of FASER's decay volume. 
As illustrated in Fig.~\ref{fig:zdc_det}, we placed decay volumes of the mentioned dimensions before the ZDCs on both sides of the IP.
Using this setup, we computed the efficiency of detecting $\phi$ and HNL in the ZDC and compared it to FASER. As shown in Fig.~\ref{fig:zdc_vs_faser}, the ZDC exhibits better efficiency than FASER. 
To further investigate, we simulated a smaller sized cuboid decay volume for the ZDC, denoted as ZDC B, with dimensions 1.5 meters in length, 96 mm in height, and 300 mm in width. This decay volume is more comparable to FASER's decay volume size. Even with this smaller decay volume, the ZDC still demonstrates better efficiency than FASER as shown in Fig~\ref{fig:zdc_vs_faser}.
This improved efficiency of the ZDC can be attributed to the ZDC's proximity to the IP, its larger decay volume, and its coverage of both positive and negative $\eta$ regions, whereas FASER only covers one side. Finally, while our study focused on the CMS ZDCs, it is likely that the ZDCs of other LHC experiments~\cite{ATLAS_ZDC,ALICE_ZDC} could possess similar capabilities for LLP searches, provided they are modified and adapted accordingly.

\begin{figure}[hbt!]
\includegraphics[width=0.9\linewidth]{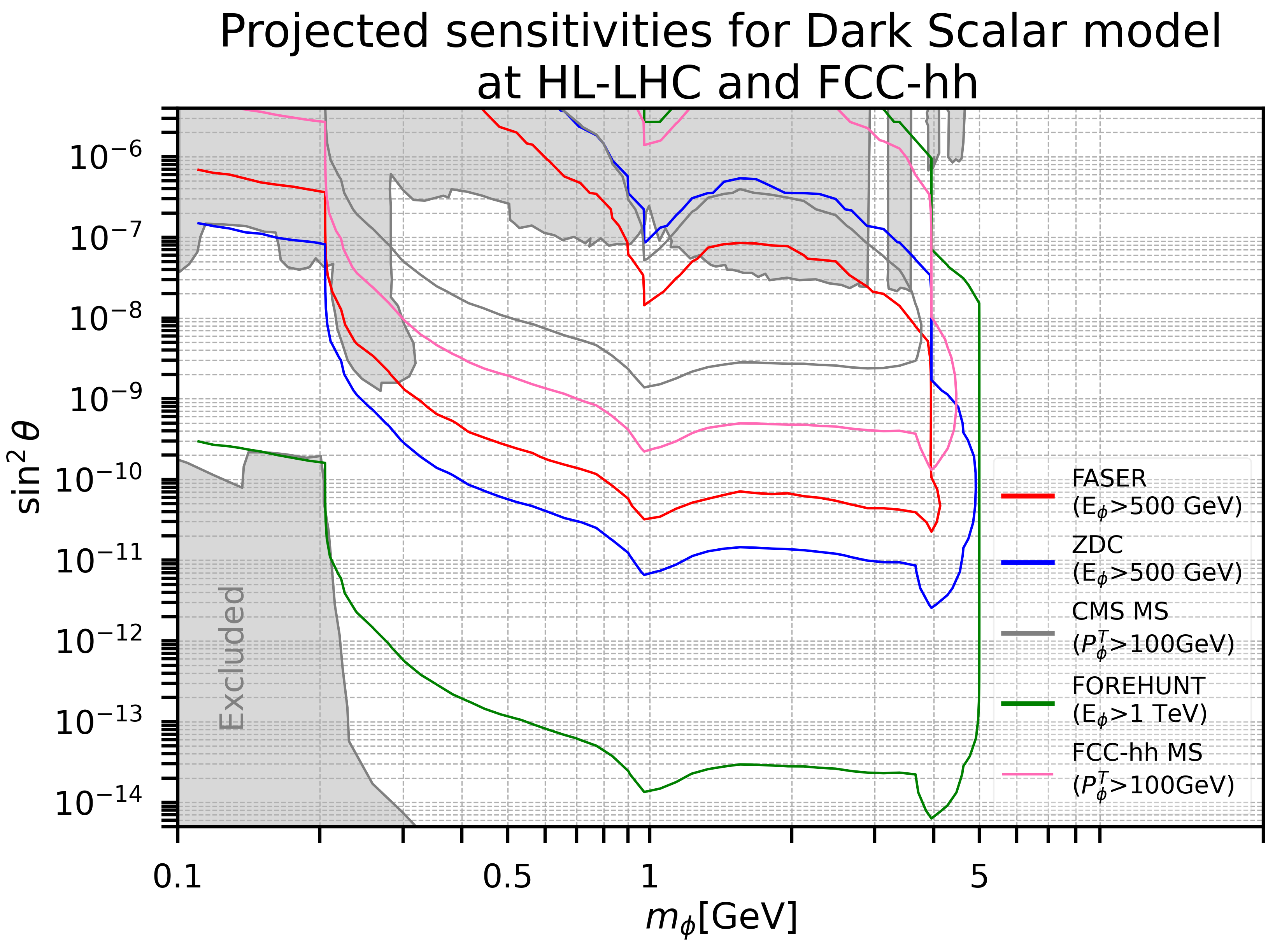}\\
\includegraphics[width=0.9\linewidth]{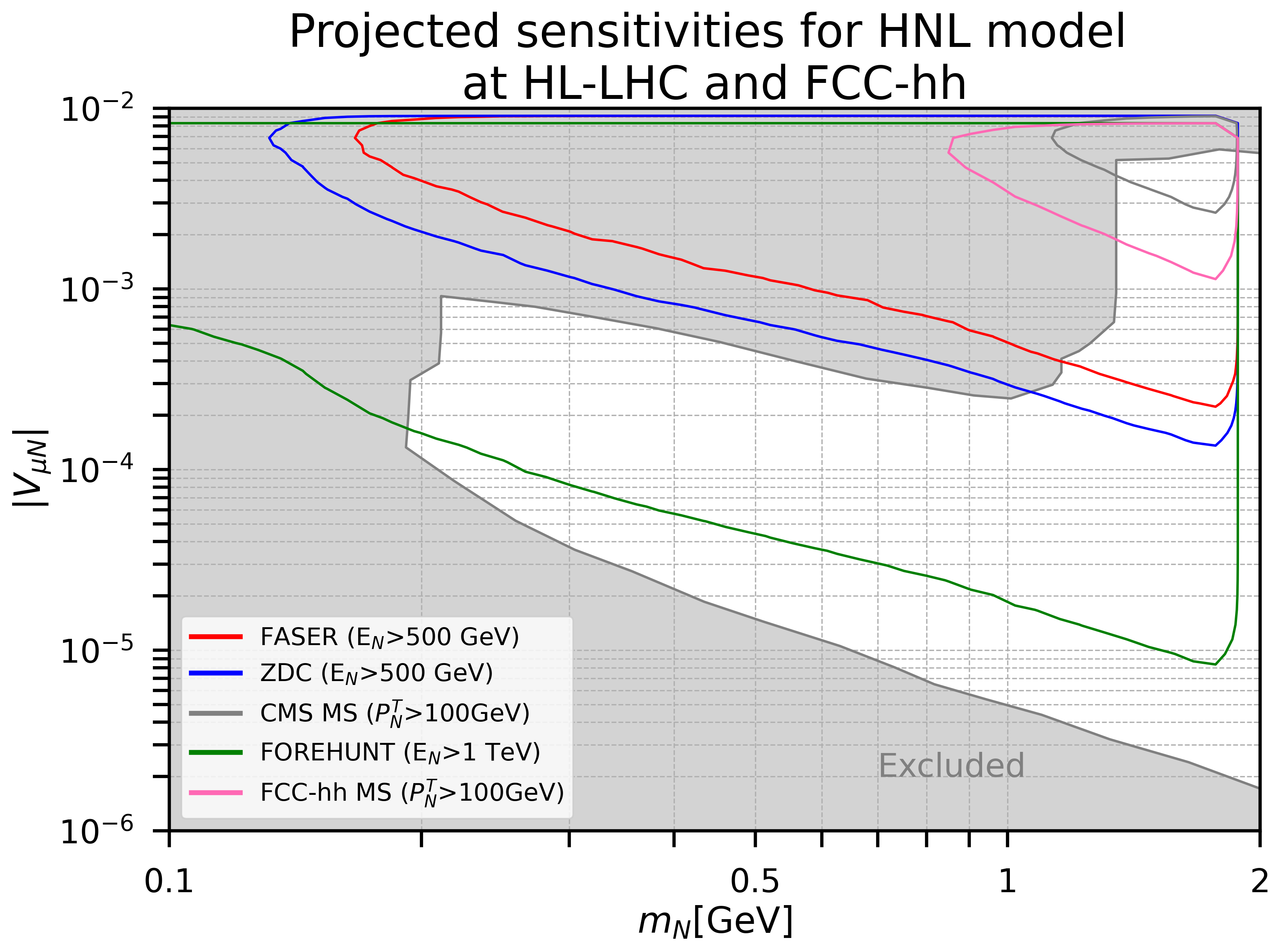}
\caption{\label{fig:model_exclusions} Projected exclusion limits for the dark-scalar model ({\it top}) and the HNL model ({\it bottom}) at the HL-LHC (CMS muon spectrometer, FASER, and ZDC) and at FCC-hh (FCC-hh muon spectrometer and FOREHUNT), including all kinematic selections indicated in the plot labels. We show the present constraints in {\it grey}, taken from Ref.\,\cite{Feng:2022inv}.}
\end{figure}

In Fig.\,\ref{fig:model_exclusions}, we present the projected exclusion limits for the dark-scalar and HNL models obtained from the CMS muon spectrometer, FASER and ZDC at the HL-LHC (14~TeV, 3~ab$^{-1}$, pileup = 200), as well as from the FCC-hh muon spectrometer and FOREHUNT at FCC-hh (100~TeV, 30~ab$^{-1}$, pileup = 1000), incorporating the kinematic selection cuts indicated in the plot labels. 
We also show the existing bounds on the parameter space (taken from Ref.\,\cite{Feng:2022inv}) coming from meson decay searches at NA62\,\cite{NA62:2020xlg}, E949\,\cite{BNL-E949:2009dza}, KOTO\,\cite{Egana-Ugrinovic:2019wzj}, LHCb\,\cite{LHCb:2012juf}, BaBar\,\cite{BaBar:2013npw}, Belle\,\cite{Belle:2009zue} and MicroBOONE\,\cite{MicroBooNE:2021usw}; beam-dump experiments such as the CHARM\,\cite{CHARM:1985anb}, LSND\,\cite{Foroughi-Abari:2020gju} and PS191\,\cite{Gorbunov:2021ccu}, and from the big bang nucleosynthesis (BBN) bound on the lifetime of the LLP\,\cite{Fradette:2017sdd}.
For the HNLs, the strongest bound for high mixing angles come from peak searches of $\tau$ lepton and meson decays and searches of displaced decays of the HNLs at beam dump and collider experiments, while for the lower mixing angles, the BBN bound excludes part of the parameter space.
For FASER, ZDC and FOREHUNT, the limits are derived using $B$-mesons originating from \textsc{SoftQCD} processes. In contrast, for the CMS muon spectrometer we are unable to obtain meaningful limits from \textsc{SoftQCD}, owing to the requirement of a 100--200\,GeV $p^{\text{miss}}_{T}$ threshold in the muon system for triggering LLP events~\cite{mitridate2023energetic}. Consequently, we use \textsc{HardQCD} processes with $p^{\text{miss}}_{T}$ threshold of 100\,GeV to determine the CMS-MS sensitivity. This stringent $p^{\text{miss}}_{T}$ requirement is the primary reason why the CMS-MS sensitivity is weaker compared to that of the dedicated LLP detectors. A similar approach is adopted for the FCC-hh muon spectrometer. We find that, for both benchmark models, the best projected sensitivities are obtained at the HL-LHC using the proposed upgraded ZDCs for LLP searches, and at the FCC-hh with the proposed FOREHUNT detector.


The Precision Proton Spectrometer (PPS)~\cite{CMS_detector_Run3} in CMS, comprising tracking and timing detectors, is primarily designed to measure protons escaping along the LHC beam line after interactions. Installed on both sides of the CMS IP, approximately $\pm200$\,m away in the LHC tunnel, PPS's potential for searching new physics in the high mass region during HL-LHC has been discussed in Ref.~\cite{PPS_Phase2}. However, its capabilities can also be leveraged for low mass LLP searches, particularly when the LLPs originate from $B$ and $D$ hadron decays and travel in the very forward directions. The timing detectors will play a crucial role in LLP searches, as they can distinguish between charged particles directly emitted from the IP and those resulting from LLP decays, based on their distinct timing signatures. Similarly, the ATLAS forward proton detectors~\cite{ATLAS_PPS} also offer promising opportunities for these searches.

Notably, when searching for low-mass BSM particles, any PU vertex can be as relevant as the primary vertex (PV), which is identified as the vertex with the largest $\sum p_T^2$ of its associated tracks~\cite{TDR_Phase2_CMS}. With $\mathcal{O}(100)$ PU, the PV may be less interesting than one of the PU vertices. For instance, it was pointed out in the context of CMS HLT for HL-LHC~\cite{cms_hlt_tdr_phase2} that for $30\%$ of SM $\text{H} \to \tau \tau$ events, the identified PV does not correspond to the vertex with the actual $\tau$ production. This insight, that interesting physics may occur in PU vertices, should guide the development of strategies for triggering and offline analysis for high PU environments.
In the HL-LHC era, dedicated trigger strategies, similar to the CMS $B$-parking strategy successfully employed in 2018, may prove beneficial. This approach targeted $b\bar{b}$ processes, where one $B$ hadron (tag-side) underwent semileptonic decay, while the other $B$ hadron (probe-side) could decay into any final state, including BSM LLPs. The tag-side was utilized in triggering by identifying semileptonic $B$ hadron decays featuring at least one displaced muon with a $p_T$ threshold as low as 7~GeV. In 2018, this strategy enabled CMS to accumulate a large, $B$ hadron enriched data sample, comprising $\mathcal{O}(10^{10})$ $b\bar{b}$ events at $\sqrt{s}=13$ TeV. 

We have investigated the prospects and feasibility of implementing $B$-parking L1 triggers in the context of HL-LHC, assuming that HLT can be implemented downstream if the L1 trigger is feasible. To validate our setup, we computed the 2018 $B$-parking L1 rate and found reasonable agreement with CMS numbers as described in Appendix~\ref{appC_Bparking}. 
The successful validation of our setup gives us the confidence to move forward with $B$-parking L1 rate estimation for HL-LHC.
At a PU of 140, a single muon L1 trigger with $p_T > 7$ (12)~GeV will yield a rate of 104 (18)~kHz from $b\bar{b}$ events. The actual rate may be higher, depending on the purity of the trigger. In the HL-LHC era, the CMS experiment expects to achieve significantly improved L1 muon purity compared to 2018. This improvement is attributed to the ability to run tracking algorithms at L1, which is currently not possible. Further details on the single muon $B$-parking L1 trigger rate in HL-LHC and FCC-hh can be found in Appendix~\ref{appC_Bparking}.

As PU increases, the feasibility of a single muon $B$-parking L1 trigger becomes increasingly challenging. At high PU, multiple $B$ hadrons are produced in different PU vertices at every bunch-crossing, making the L1 trigger rate overwhelmingly large. This issue is particularly pronounced in the FCC-hh scenario with PU$=$500, where $\sim40$ $B$ hadrons are produced per bunch-crossing, rendering the low-$p_T$ single muon L1 trigger rate unsustainable, as shown in Table~\ref{tab:bparking_HL_LHC_FCC_hh} in Appendix~\ref{appC_Bparking}. In such scenarios, increasing the $p_T$ thresholds or accepting one out of $n$ events (i.e, using a trigger prescale factor~\cite{CMS_muon_trig} of $n$) may not be the best solutions. Instead, alternative strategies can be employed, such as:
 
-- Require two muons instead of one, which may compromise the tag-and-probe setup for some events. However, for a significant fraction of events, the two muons will originate from different PU vertices, allowing for unbiased probe decays.
     
-- For extremely high PU, do not allocate any L1 or HLT rate budget for $B$-parking. Instead, utilize all events passing the HLT for searching LLP coming from the decay of $B$ (or $D$) hadron. Given the high multiplicity of $b\bar{b}$ pairs per bunch crossing, dedicated $B$-parking triggers become redundant, and all saved events will contain $B$ hadrons. This strategy may already be employed at HL-LHC where every bunch-crossing will have 2--4 $B$ hadrons. 

-- By not allocating rate to $B$-parking, we can free up rate budget, which can then be reassigned to dedicated LLP parking. Although $B$ (and $D$) hadrons will be abundant in high PU scenarios, LLPs will still be extremely rare due to the potentially small branching ratio of $B$ (or $D$) decays to LLPs. Therefore, instead of prioritizing $B$-parking, we should focus on parking events featuring signatures of LLPs. For instance, we can allocate parking rates for scenarios where LLPs decay within the muon spectrometer, resulting in clustered hits~\cite{Bhattacherjee:2021rml, CMS_trig_LLP_muon_hit_1, CMS_trig_LLP_muon_hit_2, CMS_LLP_MUON_EXO_1, CMS_LLP_MUON_EXO_2}. To maximize sensitivity to light LLPs, the L1 trigger and corresponding HLT should be designed to be as loose as possible, allowing these particles to pass the trigger selection.


As a final note on $B$-parking, as discussed before, CMS collected $\mathcal{O}(10^{10})$ $b\bar{b}$ events in 2018 through $B$-parking, with PU ranging from 24 to 54. Each of these $pp$ bunch-crossing events is expected to contain 4--10 $D$ hadrons. While CMS has already performed a search for HNLs from $B$ decays in the 2018 $B$-parking data~\cite{CMS_HNL_BParking}, the $\mathcal{O}(10^{10})$ $c\bar{c}$ events within this dataset also present an exciting opportunity to search for LLPs which CMS may pursue.

Building on the HL-LHC's capabilities, a specialized high PU fill could be attempted, targeting PU values of 400--500. This approach has precedent: in October 2016, a dedicated fill achieved a PU of 100~\cite{cms2016_pu100}, significantly higher than the average PU of 27 for the rest of 2016. A similar initiative for HL-LHC could provide valuable data for searching LLPs originating from $B$ or $D$ hadron decays. 

In summary, while the pursuit of TeV-scale BSM physics with higher-energy hadron colliders is a well-motivated and crucial endeavor, it's equally important to recognize the discovery potential of light LLPs emanating from $B$ or $D$ hadron decays. 
We highlight the overlooked benefits of PU and propose a paradigm shift in how we approach high PU environments. As next-generation hadron colliders will inevitably face increased PU, we emphasize that this challenge also brings opportunities. 
By leveraging PU vertices as additional sources of $B$ and $D$ hadrons, and implementing strategic upgrades to forward detectors like ZDC, we can unlock the potential for groundbreaking discoveries.
A study of the feasibility of performing LLP searches at the ZDC considering all possible backgrounds would be beneficial. 

\vskip 5 pt
\textbf{\textit{Acknowledgments}} -- 
BB and SM thank Sunanda Banerjee for fruitful discussions. 
BB acknowledges the MATRICS Grant (MTR/2022/000264) of the Science and Engineering Research Board (SERB), Government of India. The work of BB is also supported by the Core Research Grant CRG/2022/001922 of the Science and Engineering Research Board (SERB), Government of India. BB and AS are grateful
to the Center for High Energy Physics, Indian Institute of Science, for the cluster facility. 
AK would like to acknowledge IIT Kanpur for providing financial support through Institute fellowship.
The work of SM is supported by an initiation grant \texttt{(IITK/PHY/2023282)} received from IIT Kanpur. 

\bibliographystyle{utphys}

\providecommand{\href}[2]{#2}\begingroup\raggedright\endgroup

\section{Supplementary Material} 
\subsection{Appendix A} \label{appA_pt_pz_plot}
\begin{figure}[h]
\includegraphics[scale=0.22]{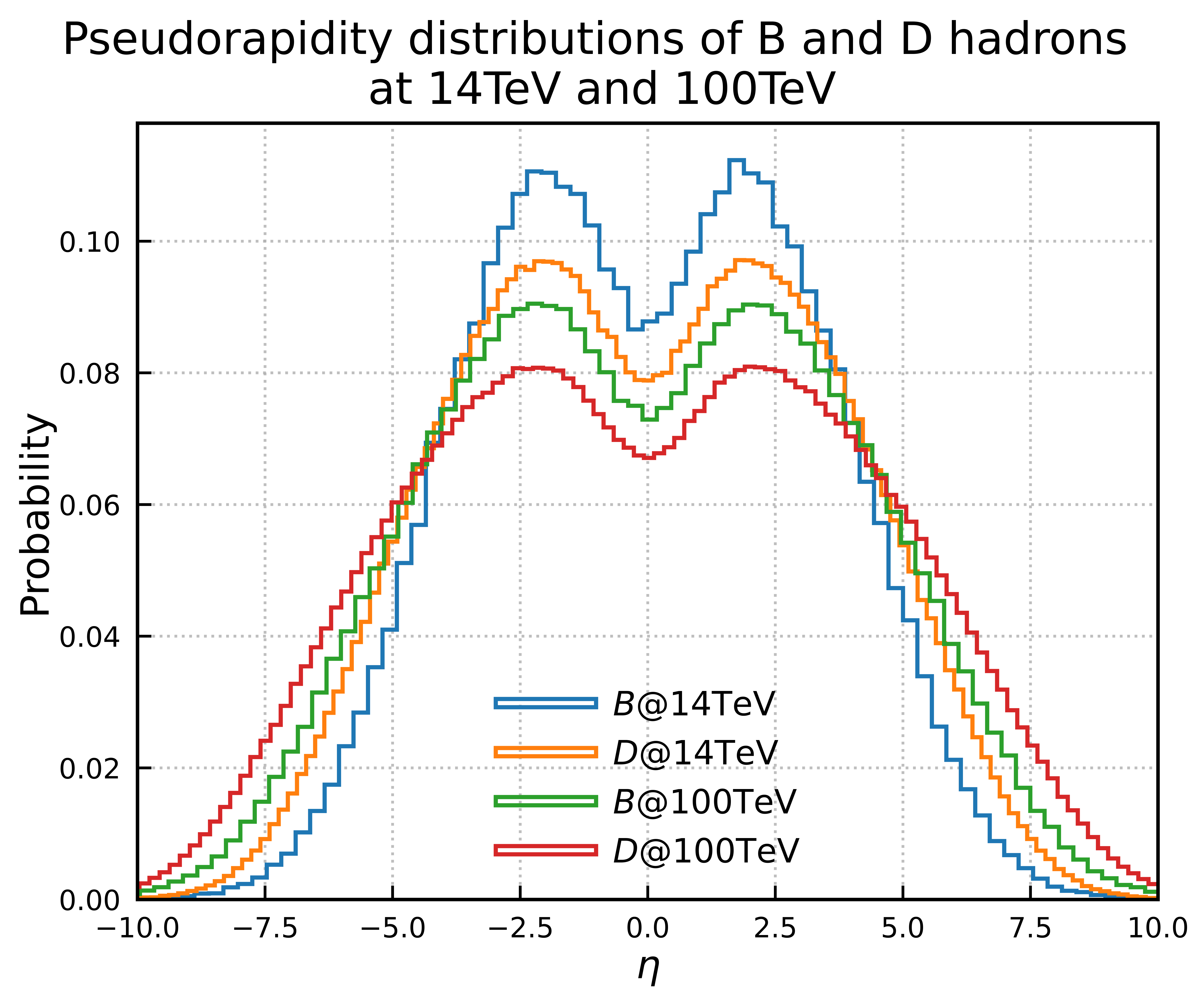}
\caption{\label{fig:Eta_had_both} Pseudorapidity distribution of $B$ and $D$ hadrons at $\sqrt{s}=100$~TeV and $\sqrt{s}=14$~TeV. In Appendix~\ref{appA_pt_pz_plot} we also provide distributions of transverse momentum ($p_T$) and longitudinal momentum ($p_z$) for 14~TeV and 100~TeV.}
\end{figure}

\begin{figure}[h]
\includegraphics[scale=0.22]{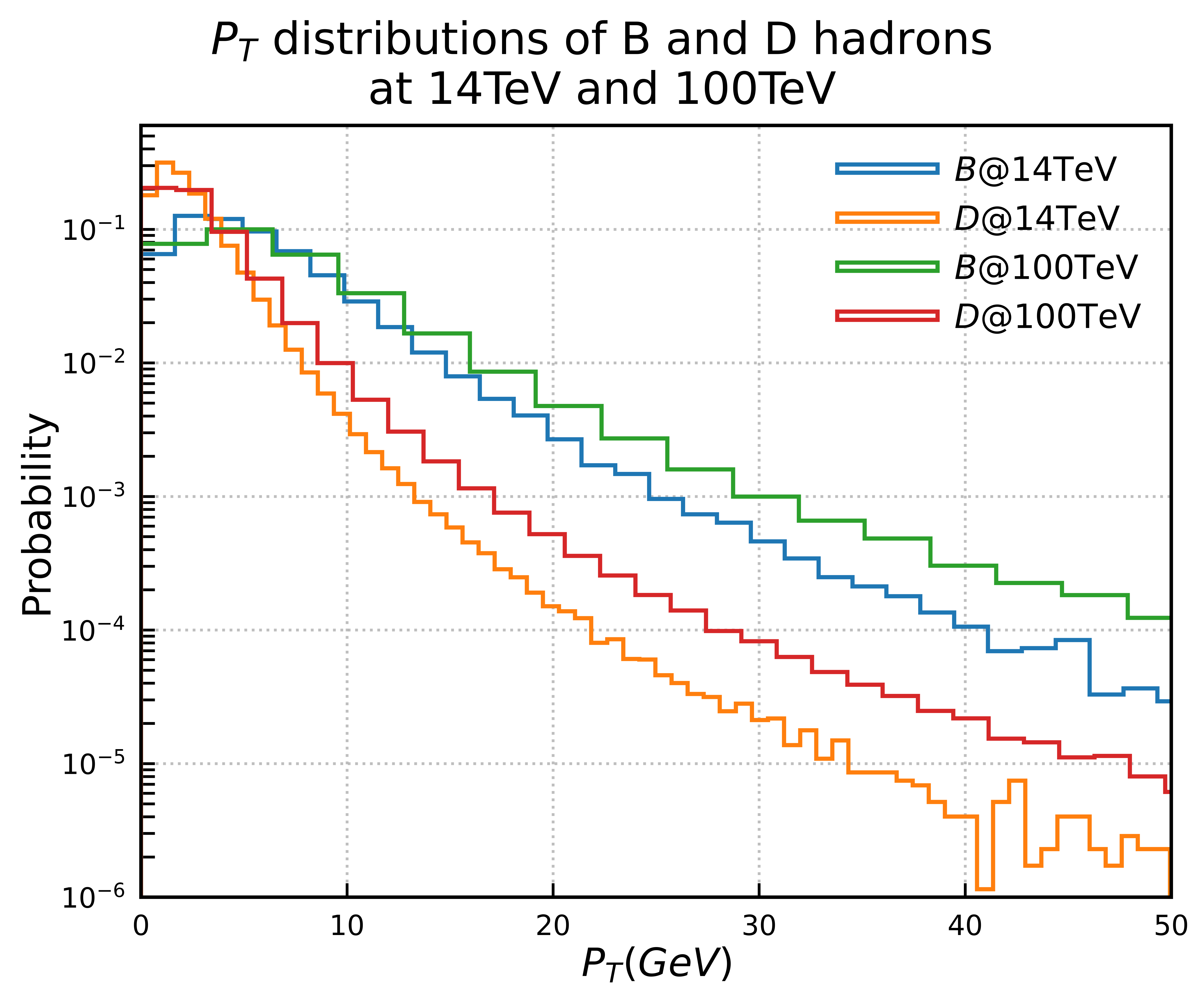}
\caption{\label{fig:pz_had_both} Longitudinal momentum ($p_z$) distribution of $B$ and $D$ hadrons at $\sqrt{s}=100$~TeV and $\sqrt{s}=14$~TeV.}
\end{figure}

\begin{figure}[h]
\includegraphics[scale=0.22]{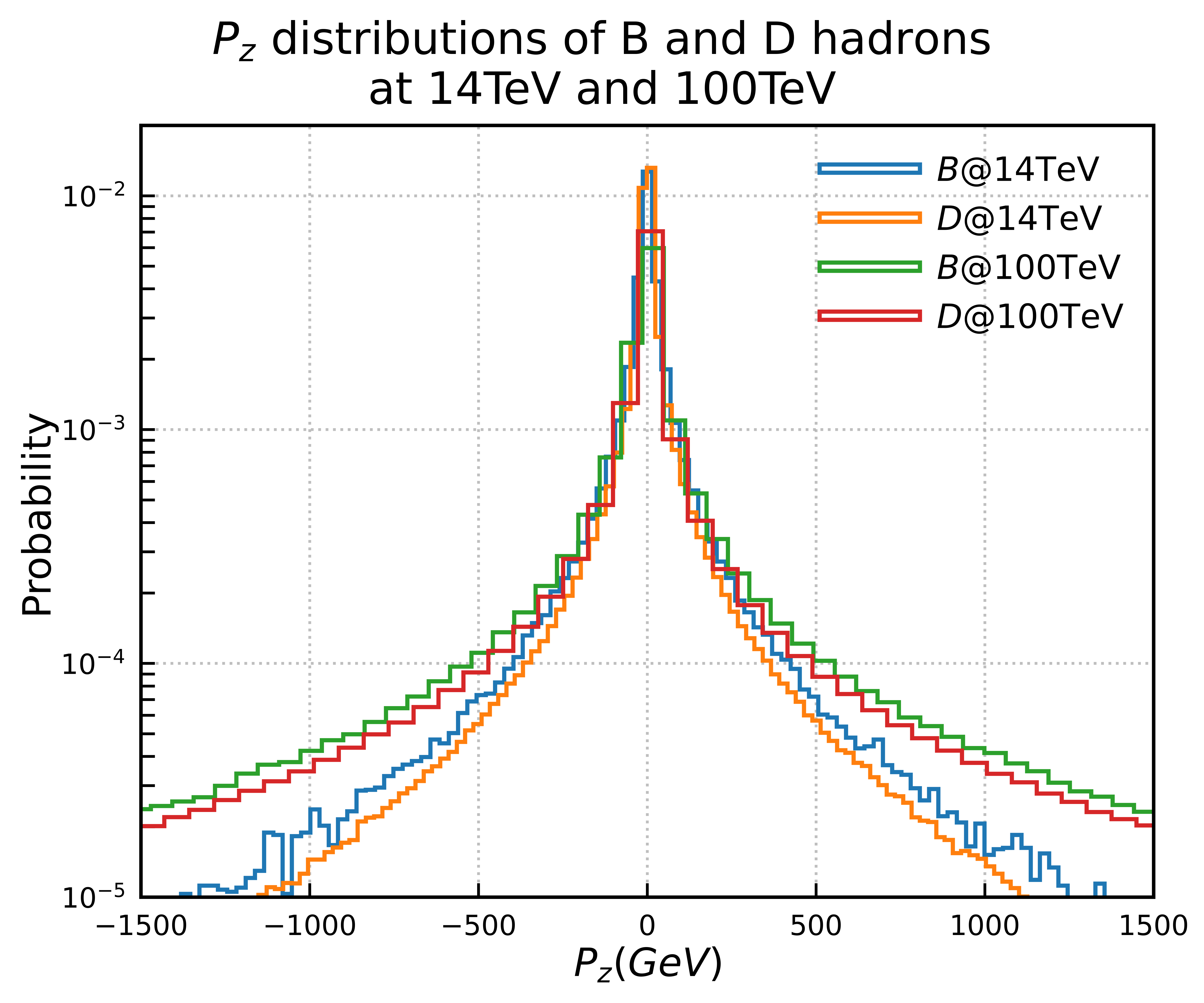}
\caption{\label{fig:pt_had_both} Transverse momentum ($p_T$) distribution of $B$ and $D$ hadrons at $\sqrt{s}=100$~TeV and $\sqrt{s}=14$~TeV.}
\end{figure}

Figures~\ref{fig:Eta_had_both}, \ref{fig:pz_had_both} and Figure~\ref{fig:pt_had_both} respectively show the distributions of the pseudorapidity ($\eta$), longitudinal momentum ($p_z$) and transverse momentum ($p_T$) of $B$ and $D$ hadrons produced in soft QCD collisions at $\sqrt{s}=14$~TeV and $\sqrt{s}=100$~TeV.
%
%
We found that in $\sqrt{s}=14$ (100)~TeV, average $p_z$ of $B$ hadrons is $\sim$115 (450)~GeV, while average $p_T$ is $\sim$ 6 (7)~GeV. Similarly, for $D$ hadrons the average $p_z$ is $\sim$87 (350)~GeV, while average $p_T$ is $\sim$ 2 (3)~GeV. These findings underscore the importance of dedicated forward detectors in enhancing the sensitivity of searches for LLPs originating from the decay of these hadrons, as a substantial fraction of them are emitted in the forward region.

\vskip 10 pt
\subsection{Appendix B} \label{appB_fms_zdc_faser_eff}

\begin{figure}[h]
\includegraphics[scale=0.20]{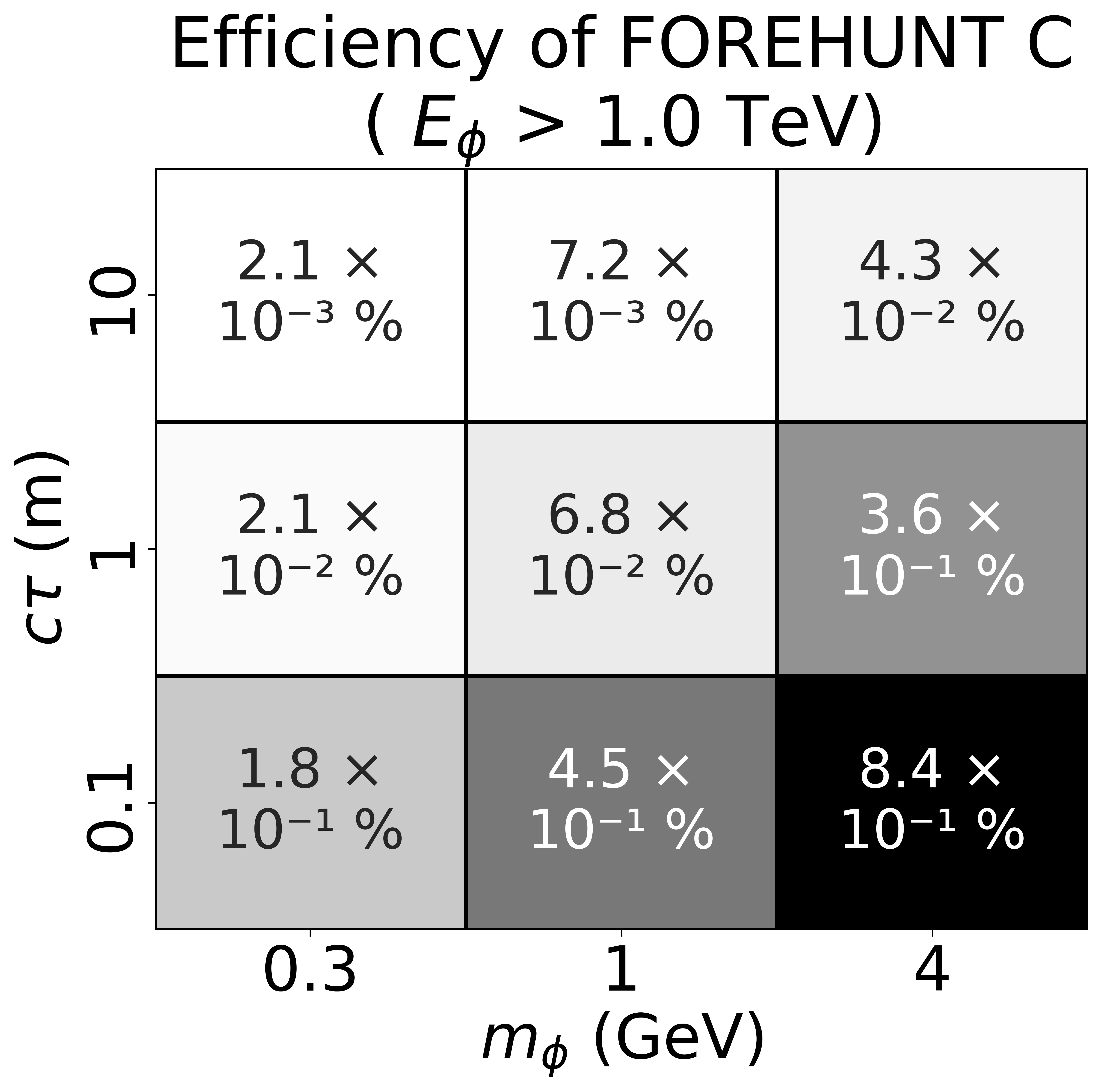}
\hskip 5 pt
\includegraphics[scale=0.20]{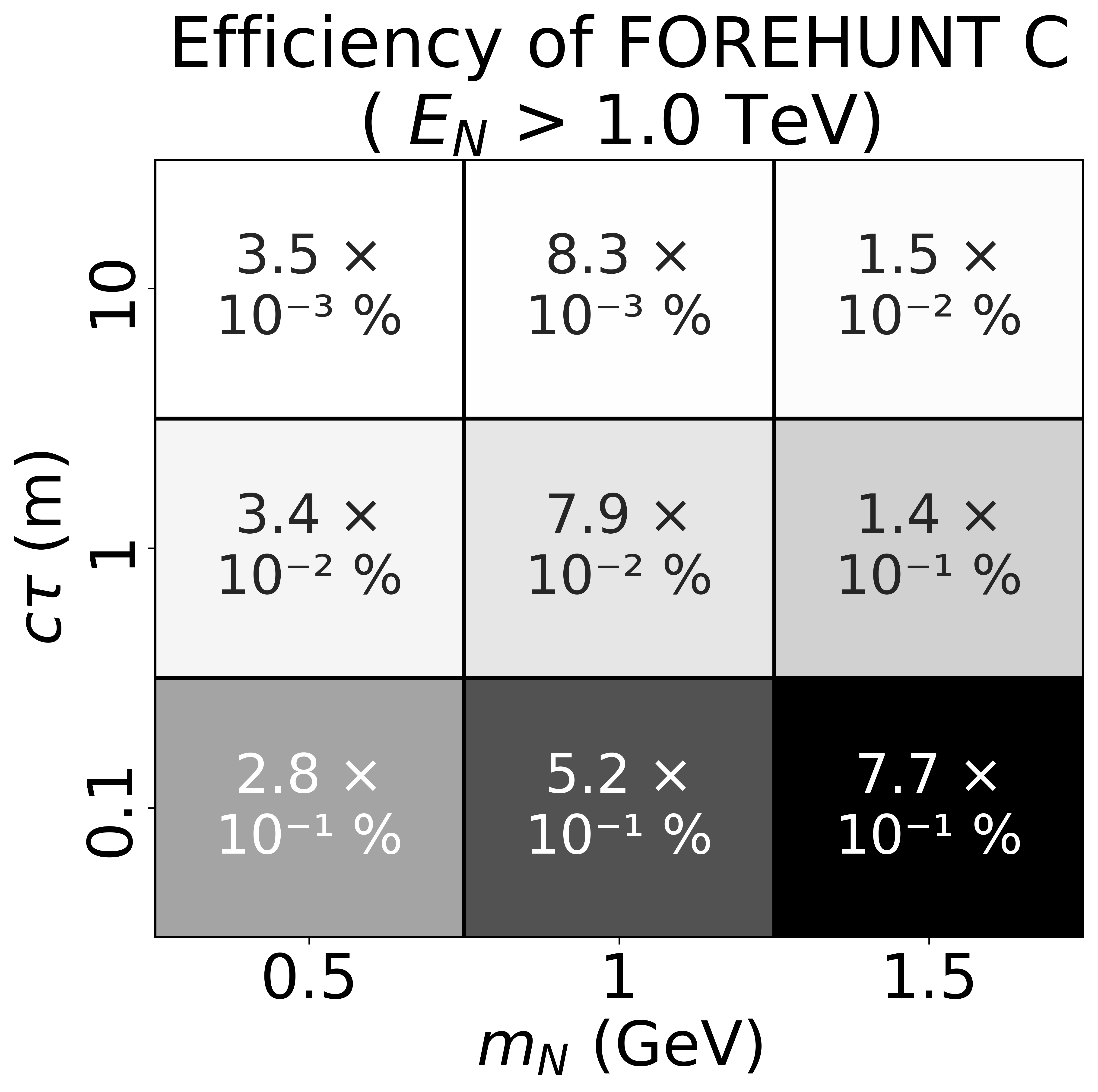}
\caption{\label{fig:forehunt_C_eff_E_1TeV} Efficiency of detecting $\phi$ ({\it left}) and $N$ ({\it right}) for the cylindrical FOREHUNT detector with configuration C where the energy of the LLP is required to be $>$1~TeV. The detector surrounds the beampipe, with its decay volume defined to exclude the FCC-hh beampipe (assumed to have a radius of 10~mm) for the purpose of this study. Further details on the various FOREHUNT configurations can be found in Ref.~\cite{forehunt_paper}.}
\end{figure}
\begin{figure}[hbt!]
\includegraphics[scale=0.20]{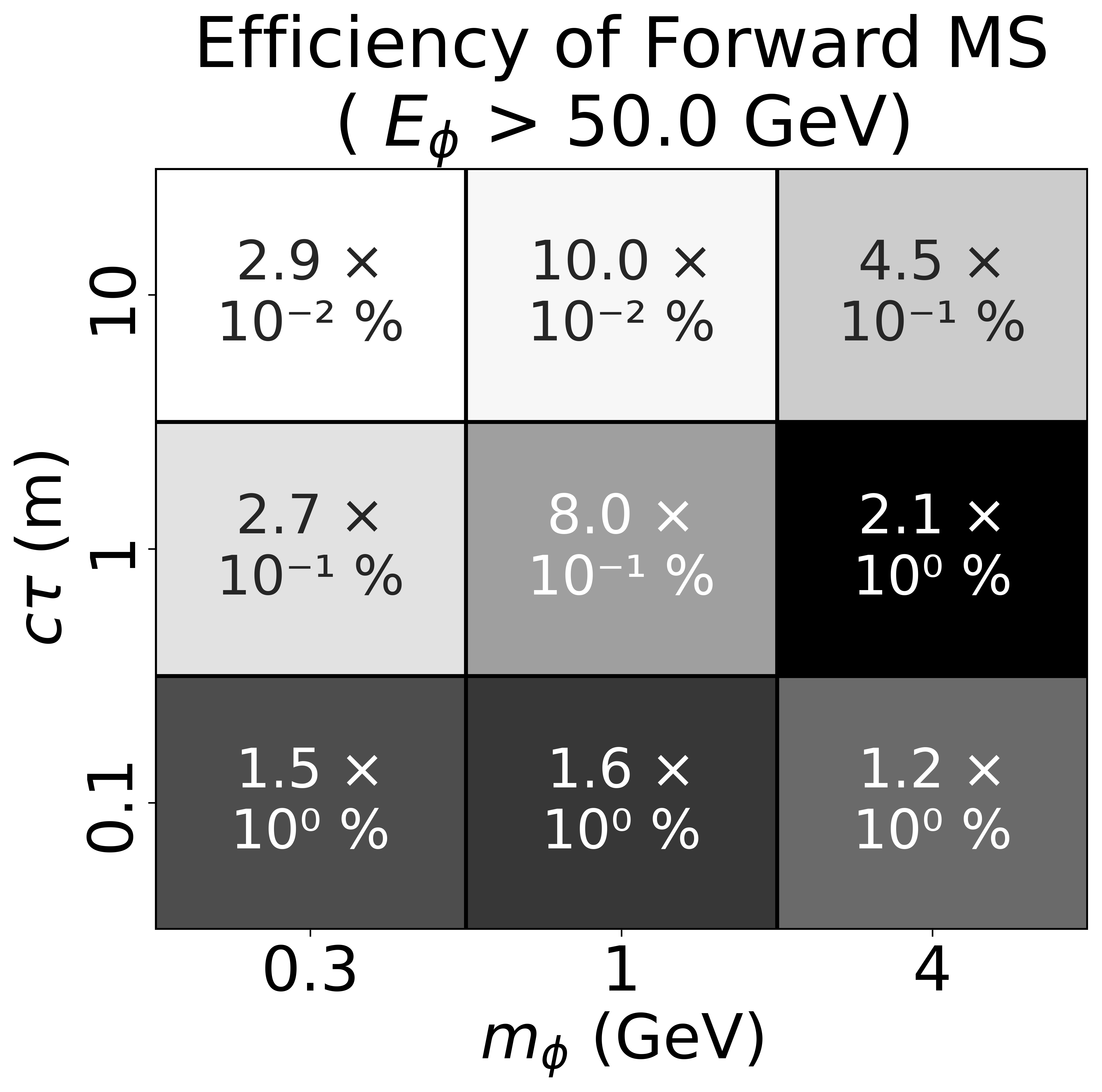}
\hskip 5 pt
\includegraphics[scale=0.20]{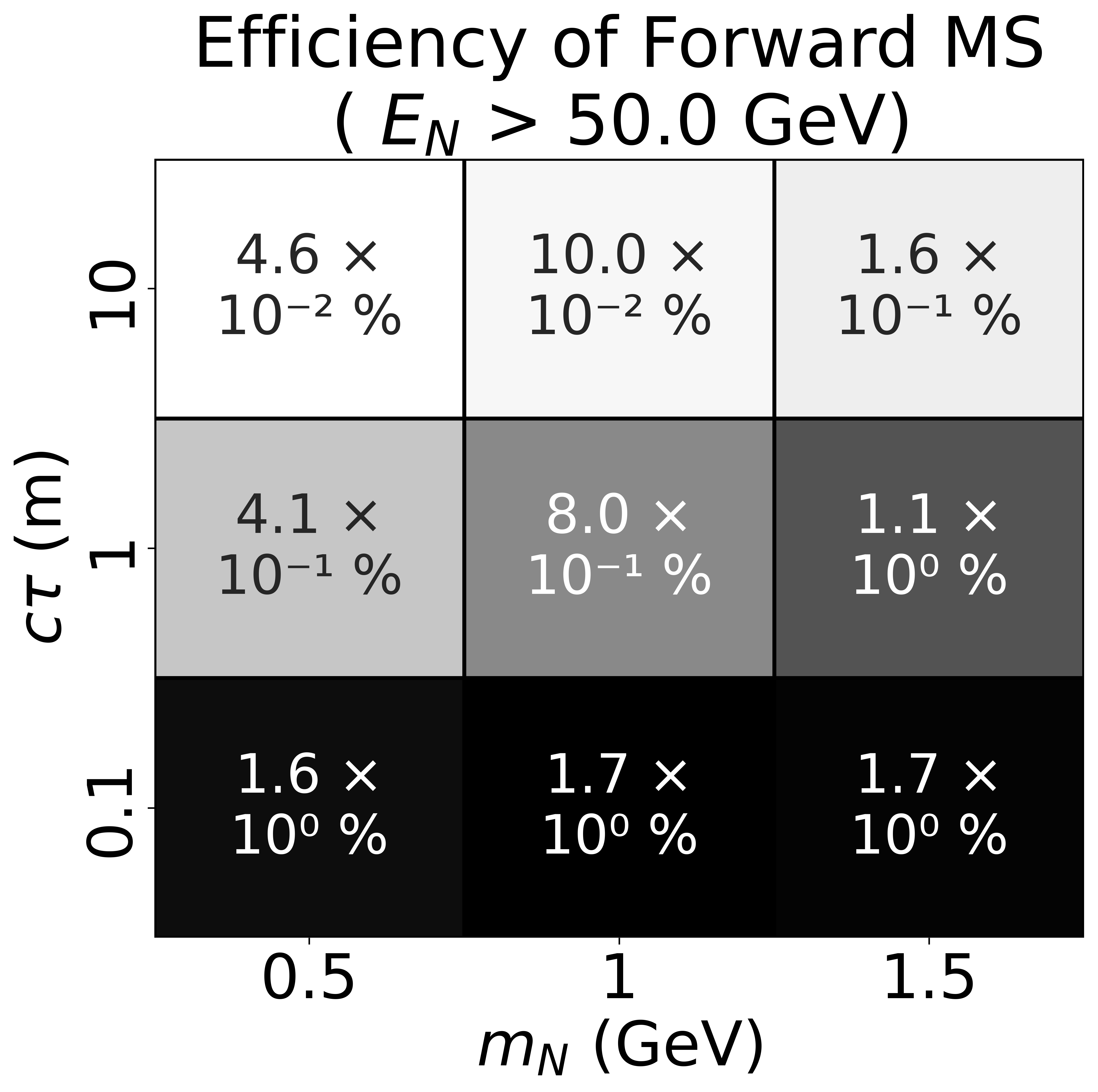}
\caption{\label{fig:FMS_eff_E_50GeV} Efficiency of detecting $\phi$ ({\it left}) and $N$ ({\it right}) for the forward muon spectrometer (FMS), which is a part of the main detector of FCC-hh. The energy of the LLP is required to be $>$50~GeV. }
\end{figure}
\begin{figure}[hbt!]
\includegraphics[scale=0.20]{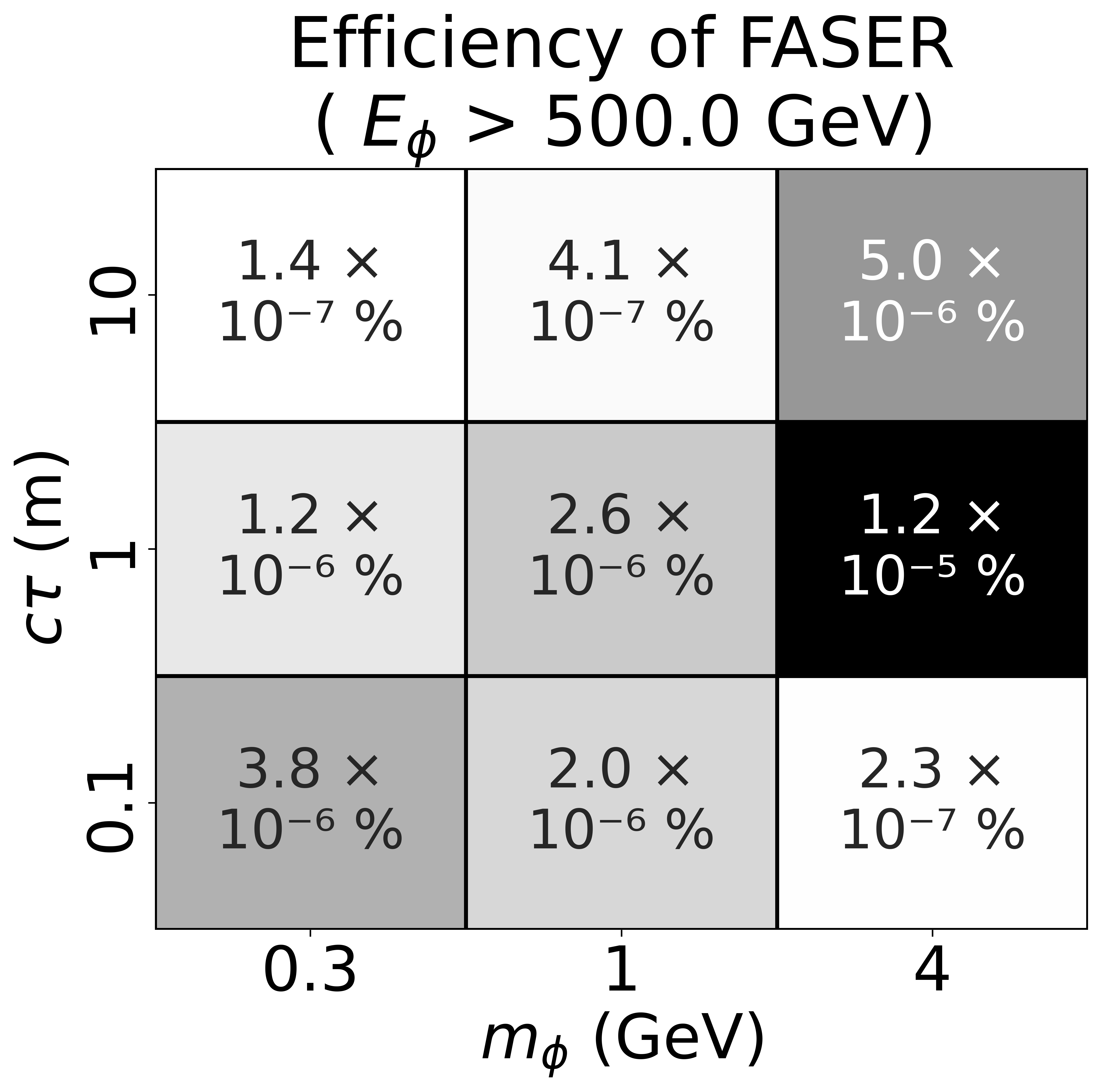}
\hskip 5 pt
\includegraphics[scale=0.20]{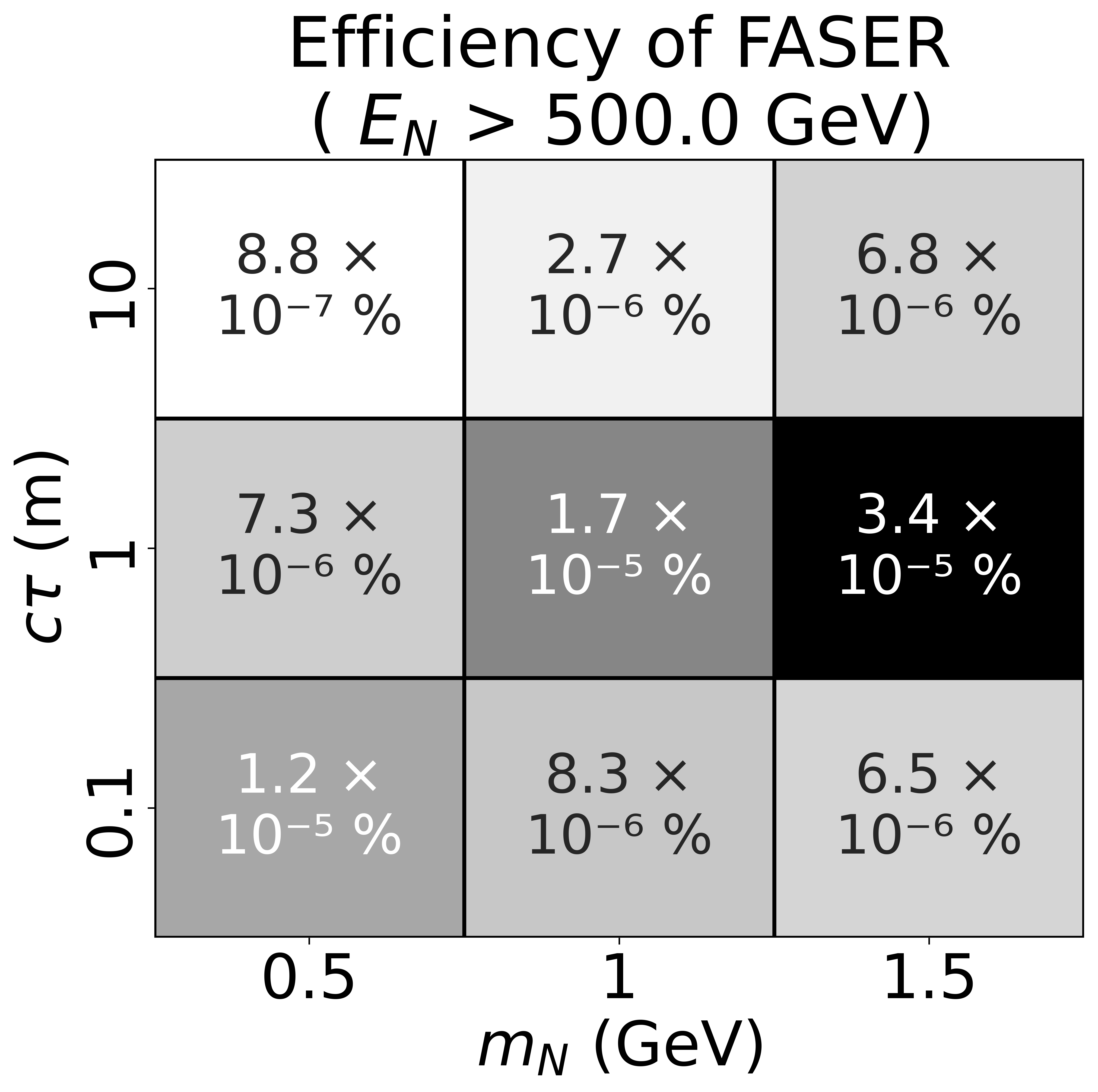}
\caption{\label{fig:FASER_eff_E_500GeV} Efficiency of detecting $\phi$ ({\it left}) and $N$ ({\it right}) for the FASER detector, which is a part of the main detector of FCC-hh. The energy of the LLP is required to be $>$500~GeV. }
\end{figure}
\begin{figure}[hbt!]
\includegraphics[scale=0.20]{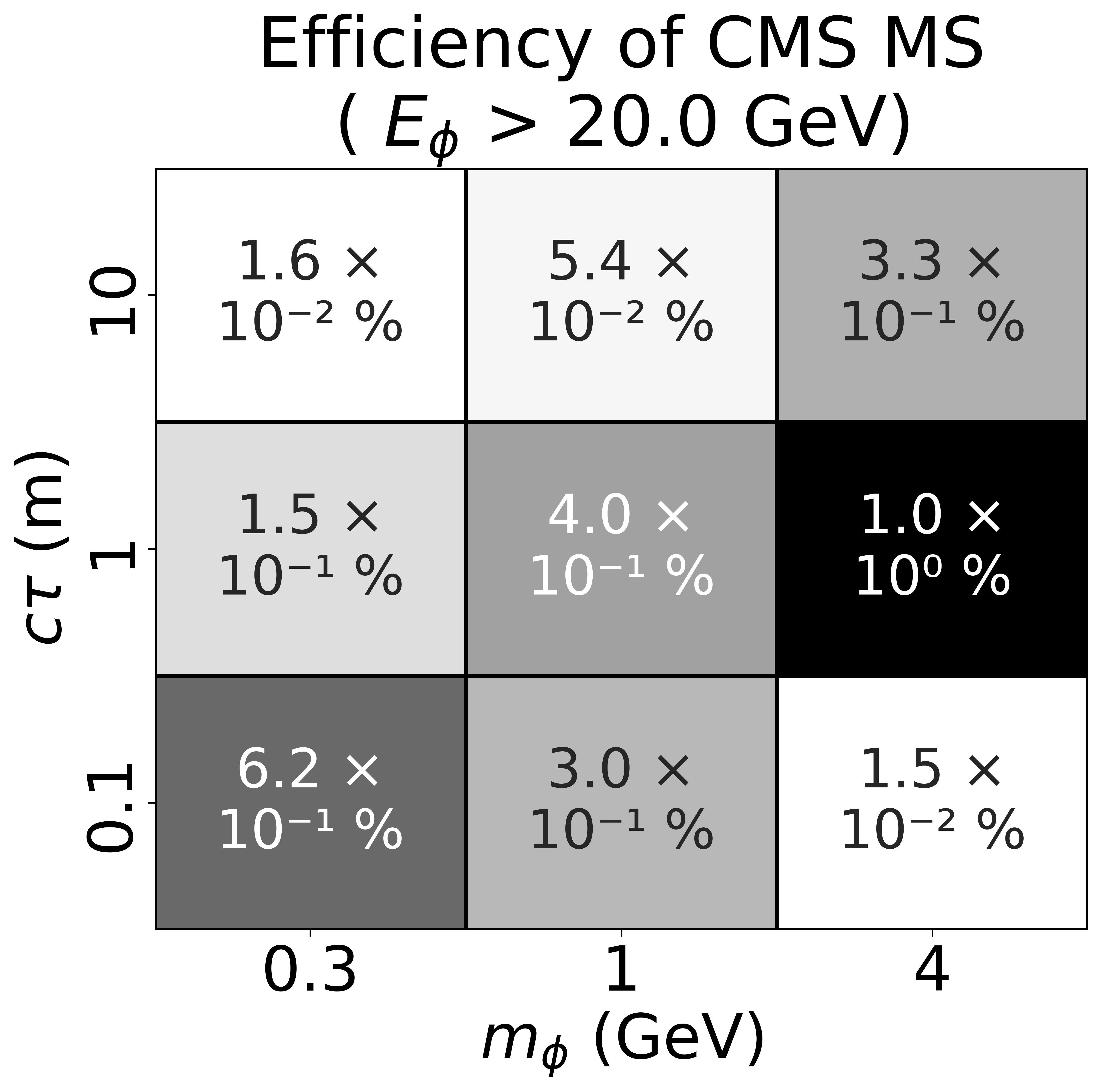}
\hskip 5 pt
\includegraphics[scale=0.20]{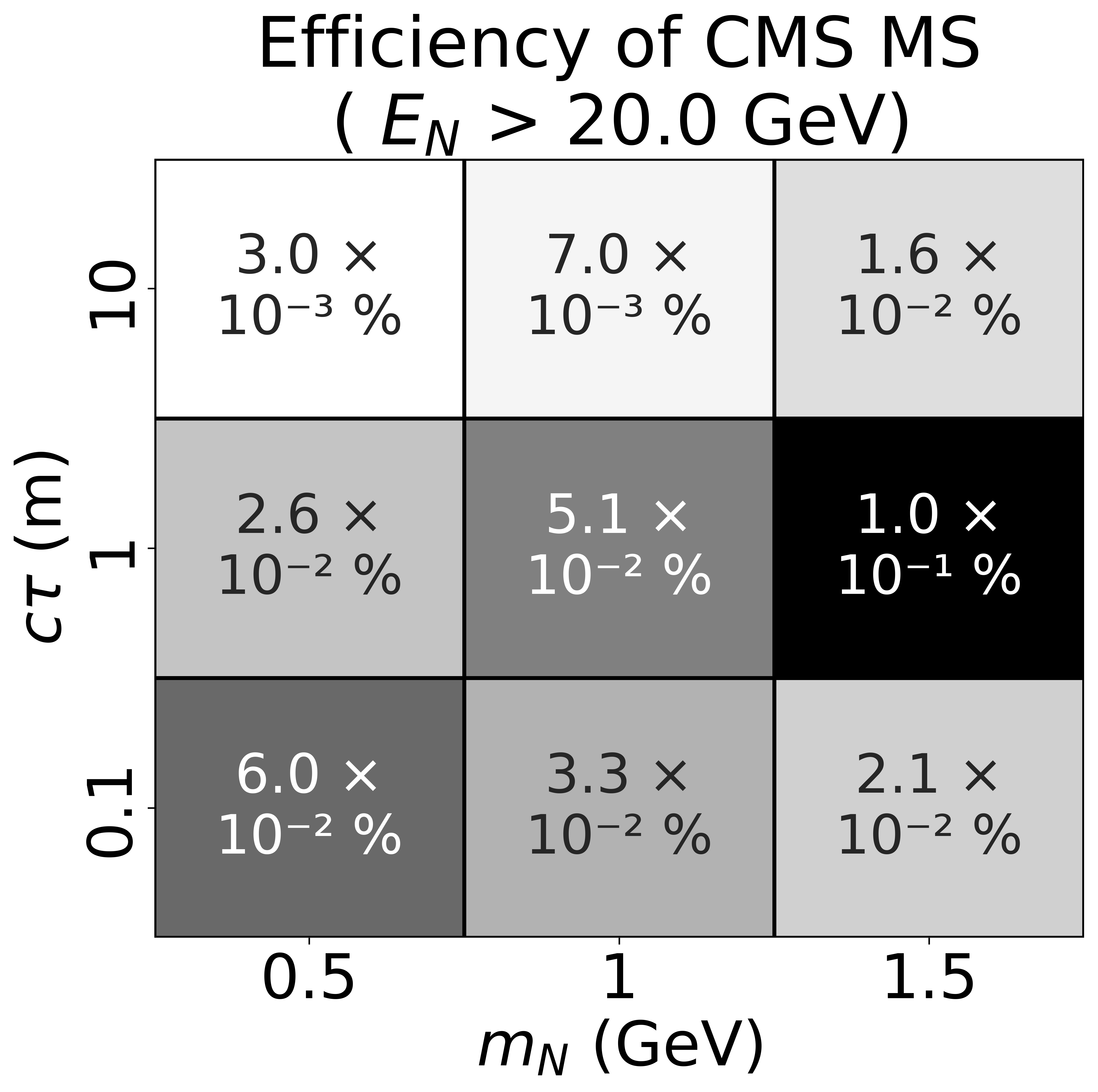}
\caption{\label{fig:CMS_eff_E_20GeV} Efficiency of detecting $\phi$ ({\it left}) and $N$ ({\it right}) in the CMS muon spectrometer (MS) in HL-LHC, where the energy of the LLP is required to be $>$20~GeV.}
\end{figure}

Figures~\ref{fig:forehunt_C_eff_E_1TeV}, \ref{fig:FMS_eff_E_50GeV}, \ref{fig:FASER_eff_E_500GeV} and \ref{fig:CMS_eff_E_20GeV} show the detection efficiencies in 2D plane of mass and $c\tau$ for the two benchmark scenarios of LLPs, $\phi$ ({\it left}) and $N$ ({\it right}), respectively in the FOREHUNT-C, the forward muon spectrometer (FMS) of the FCC-hh main detector, the FASER detector and the CMS muon spectrometer. The required energy thresholds on the LLPs for the different detectors are mentioned in the respective figures.


\vskip 10 pt
\subsection{Appendix C} \label{appC_Bparking} 
To validate our L1 trigger rate measuring setup, we computed the 2018 $B$-parking L1 rate assuming same values of PU as reported by CMS. 
For muons with $p_T > 7$ (12)~GeV and $|\eta|<1.5$, we obtained an L1 rate from $b\bar{b}$ events of 18.1 (5.8)~kHz. In comparison, CMS reported peak L1 rates of 53~kHz and 20~kHz for the 7~GeV and 12~GeV $p_T$ cuts, respectively, which include a substantial background~\cite{CMS_scouting_parking}. Accounting for the average L1 trigger purity of 0.3 for $b\bar{b}$ events, the CMS L1 rates attributable to $b\bar{b}$ events are 15.9~kHz (53~kHz$\times$0.3) and 6~kHz (20~kHz$\times$0.3) for the 7~GeV and 12~GeV $p_T$ cuts. Our results match reasonably well with these values. More details are provided in Table~\ref{tab:bparking_2018}. We then estimated the L1 single-muon trigger rate from $b\bar{b}$ events under HL-LHC and FCC-hh conditions, with the results presented in Table~\ref{tab:bparking_HL_LHC_FCC_hh}. Even assuming 100\% trigger-level purity for $b\bar{b}$ events, the low-$p_T$ single-muon trigger rate will become increasingly unsustainable with rising PU. For example, at $\text{PU}=500$, single-muon L1 trigger rate from $b\bar{b}$ events can be as high as 535~kHz for muon $p_T>12$~GeV. This is due to the fact that, at high PU, multiple $b\bar{b}$ pairs will be produced at every $pp$ bunch-crossing.
\begin{table}[hbt!]
\caption{\label{tab:bparking_2018}%
L1 single-muon $B$-parking trigger rates in 2018 $pp$ run at the LHC at $\sqrt{s}=13$~TeV.
}
\begin{ruledtabular}
\begin{tabular}{ccccc}
 & 
\textrm{$\mu$ $p_T$} & 
\textrm{Peak}& 
\textrm{L1 rate}& 
\textrm{L1 rate}\\ 
\textrm{PU} & 
\textrm{threshold} & 
\textrm{L1 rate}& 
\textrm{from $b\bar{b}$}& 
\textrm{from $b\bar{b}$}\\ 
 & 
\textrm{[GeV]} & 
\textrm{(CMS)}& 
\textrm{(CMS)}& 
\textrm{(our setup)}\\ 
 & 
 & 
\textrm{[kHz]}&
\textrm{[kHz]}& 
\textrm{[kHz]}\\ 
\colrule
24.3 & 7   & 53   &  15.9  & 18.1\\
29.7 & 8   & 43   &  12.9  & 14.9\\
35.1 & 9   & 32   &  9.6   & 11.7\\
42.8 & 10  & 30   &  9.0   & 10.0\\
45.9 & 12  & 20   &  6.0   & 5.8\\
\end{tabular}
\end{ruledtabular}
\end{table}
%
\begin{table}[hbt!]
\caption{\label{tab:bparking_HL_LHC_FCC_hh}%
L1 single-muon $B$-parking trigger rates in HL-LHC and FCC-hh conditions with $\text{PU}=140$ and $\text{PU}=500$ respectively.
}
\begin{ruledtabular}
\begin{tabular}{ccc}
\textrm{$\mu$ $p_T$} & 
\textrm{HL-LHC L1 rate} & 
\textrm{FCC-hh L1 rate}\\ 
\textrm{threshold} & 
\textrm{from $b\bar{b}$ [kHz]} & 
\textrm{from $b\bar{b}$ [kHz]}\\ 
\textrm{[GeV]} & 
\textrm{$\sqrt{s}=14$~TeV} & 
\textrm{$\sqrt{s}=100$~TeV}\\ 
& 
\textrm{$\text{PU}=140$} & 
\textrm{$\text{PU}=500$}\\ 
\colrule
5 & 266 & 5299\\
7 &  104 & 2451\\
10 & 33 & 926\\
12 & 18 & 535\\
14 & 10 & 328\\
16 & 6 & 212\\

\end{tabular}
\end{ruledtabular}
\end{table}

\end{document}